\documentclass[prl,showpacs,twocolumn]{revtex4-2}

\usepackage{graphicx}
\usepackage{amsmath}
\usepackage{slashed}
\usepackage{feynmp}

\begin{document}
\title{Model of Flavors}
\author{Ji\v r\'{\i} Ho\v sek}
\email{hosek@ujf.cas.cz}\affiliation
{Senior Physicist Emeritus, Department of Theoretical Physics,\\
Nuclear Physics Institute, Czech Acad. Sci., \v Re\v z 292, 250 68 Husinec, Czech Republic}

\begin{abstract}
\noindent The BCS-motivated idea of Weinberg and Salam on dynamical EW symmetry breaking is revived: The Higgs sector of the EW
gauge model of three fermion flavors (families) is replaced with the chiral gauge $SU(3)_f$ quantum flavor dynamics (QFD)
strongly coupled at a huge scale $\Lambda$. I. With all chiral fermions in flavor triplets the anomaly freedom demands
the welcome BSM extension of the SM fermion sector by one EW-sterile neutrino $\nu_R$ per flavor. II. The QFD distinguishes flavors
by generating at strong coupling the chirality-prohibited (i.e. calculable) fermion masses: Three different Majorana
masses $M_f \sim \Lambda$ of $\nu_R$, and three different, arguably small Dirac masses $m_f$ of SM fermions {\it degenerate for all
species in each flavor}. 1. Complete spontaneous breakdown of $SU(3)_f \times U(1)$ by $M_f$ implies:
(i) All flavor gluons acquire self-consistently masses $\sim M_f$. (ii) There is the $\nu_R$-composite pseudo-NG Majoron. (iii) There are three
very heavy $0^{+}$ $\nu_R$-composite Higgs bosons. 2. Spontaneous breakdown of the EW $SU(2)_L \times U(1)_Y$ symmetry
to $U(1)_{em}$ by $m_f$, in sharp contrast with the Higgs mechanism, implies: (i) The $W$ and $Z$ bosons acquire masses
$\sim g(\sum m^2_f)^{1/2}$ and $\sim (g^2+g^{'2})^{1/2}(\sum m^2_f)^{1/2}$ respectively, defining the effective Fermi scale $v=246 \rm GeV$.
(ii) There are three SM-fermion-composite $0^{+}$ Higgs bosons $h_f$ at this scale. III. The UV-finite EW dynamical
perturbation theory of Pagels and Stokar splits the flavor-degenerate masses of SM-fermions by their electric charges
and the ratios $m_f/m_{W,Z}$. Six Majorana neutrino masses are calculable by seasaw.
\end{abstract}

\maketitle

\section{I. Introduction}
Hard lepton and quark masses of the SM break the chiral gauge EW $SU(2)_L \times U(1)_Y$ symmetry down to $U(1)_{em}$.
Consequently, {\bf if} generated dynamically, by Goldstone theorem they imply three fermion-composite
'would-be' NG bosons giving rise to the softly massive $W$ and $Z$ bosons.
This was the first idea considered by Weinberg and Salam \cite{1wein1} for the spontaneous EW symmetry breaking. The alluring analogy
with the Meissner effect of the dynamically massive magnetic field in the {\it microscopic} BCS superconductor \cite{2bcs},
however, heavily falters: In BCS the electron gap responsible for spontaneous breakdown of the electromagnetic $U(1)$
gauge symmetry resulting in the Meissner effect can be generated by an arbitrarily weak force between electrons.
In sharp contrast the new dynamics generating the fermion masses and the corresponding 'would-be' NG bosons in the SM must be strong.
For the lack of any experimental evidence for such a force Weinberg and Salam rejected the idea as unrealistic, and suggested instead the weakly
coupled Higgs mechanism \cite{3higgs}: The point is that the Meissner effect is described also by the sophisticated, successful and predictive
{\it phenomenological} Ginzburg-Landau (GL) theory od superconductivity \cite{4gl}. The Higgs mechanism is nothing but its Lorentz-invariant
non-Abelian modification. Yet the fame of the Higgs mechanism is enormous: First, the brilliant proof of Gerard t'Hooft \cite{5thooft}
turned the SM with the Higgs mechanism into the full-fledged quantum field theory. Second, the glorious discovery of the Higgs boson
in 2012 at the CERN LHC \cite{6cernhiggs} confirmed the correctness of the Higgs mechanism. Despite of all this the tree-level {\it description}
of the soft fermion mass spectrum in the Higgs mechanism is phenomenological by construction.
For Steven Weinberg such a state of affairs could not be satisfactory. The {\it origin} of the observed pattern of lepton and quark masses
was the prime mystery of particle physics he wished to see solved \cite{7wein2}.

Strong appeal of the Weinberg-Salam (WS) idea is in the all-important role of fermions. In contrast, the theoretical role of fermions
in the Higgs mechanism is marginal: Even without them the $W$ and $Z$ bosons acquire soft masses by the Higgs-field condensate.
Unrealistic prototypes of the Meissner effect analogy in the EW interactions with the strongly coupled heavy Abelian gauge boson or the
corresponding four-fermion NJL interaction \cite{8njl} were elaborated in \cite{9hosek123}. Later the authors of papers \cite{10bhl}
and \cite{11miransky} suggested to identify the iterated scalar bubble in the strong-coupling NJL with
the top-quark-composite Higgs particle. Similar idea was advocated also by Yoichiro Nambu \cite{12nambu}. The idea of
spontaneous gauge symmetry breaking by the dynamical chiral symmetry breaking obviously does not need the superconductivity analogy.
The 'Abelian' papers \cite{13jjcn} contain already all necessary technical details of the dynamical gauge boson mass generation.
The crucial point, the strong force responsible for the generation of the fermion masses remains unspecified. Similarly, the
model of massive EW gauge boson masses by a strong non-Abelian dynamics with spontaneous chiral symmetry breaking is technicolor (TC) \cite{14tc}.

We realize the WS idea by the $SU(3)_f \times SU(2)_L \times U(1)_Y$ gauge-invariant Lagrangian
\begin{eqnarray*}
{\cal L}_f&=&  \bar q_L i\slashed D q_L + \bar u_R i\slashed D u_R + \bar d_R i\slashed D d_R\\
&+& \bar l_L i\slashed D l_L + \bar e_R i\slashed D e_R + \bar \nu_{R} i\slashed D \nu_{R}\\
&-&\tfrac{1}{4}F_{\mu \nu}F^{\mu \nu}(A) -\tfrac{1}{4}F_{\mu \nu}F^{\mu \nu}(B) -\tfrac{1}{4}F_{\mu \nu}F^{\mu \nu}(C)
\end{eqnarray*}
{\it in which all chiral fermions transform as flavor triplets.}
The experimental fact of three fermion flavors thus acquires the operationally well defined dynamical meaning \cite{15flavorchanging}.

The Lagrangian ${\cal L}_f$ is invariant also with respect to the global anomalous $U(1)_s$ 'sterility' phase symmetry of $\nu_R$.
The covariant derivatives include both the EW and QFD interactions; $q_L, l_L$ are the EW doublets, $u_R, d_R, e_R, \nu_R$ are the
EW singlets with the weak hyper-charges fixed by the fermion electric charges:
\begin{equation}
Q_s = (T_3)_s + \tfrac{1}{2}Y_s
\label{qs}
\end{equation}
Here $s = \nu, l, u, d$ for all three flavors, $C$ are the flavor gluons, and their gauge coupling
is $h$. By dimensional transmutation \cite{16coleman} the dimensionless $h$ turns into the theoretically arbitrary mass scale $\Lambda$.
In the present strong-coupling context the model was suggested by Tsutomu Yanagida in \cite{17yana1}. The dynamics generating
the fermion masses was coined but not specified, by Pagels and Stokar \cite{18ps} as the {\it quantum flavor dynamics (QFD)}.
The name exactly expresses our idea, and we take the liberty of using it here.

{\bf The hard fermion masses} are described by the Lorentz-invariant mass terms, the bridges in the Lagrangian between the right-handed (R)
and the left-handed (L) fermion fields. {\bf In ${\cal L}_f$ all of them are strictly prohibited by its global chiral symmetries
together with the corresponding mass counter terms:}\\
(i) The Majorana mass term of $\nu_R$
\begin{equation}
{\cal L}_{M_R} = -\tfrac{1}{2}[\bar \nu_R M (\nu_R)^{{\cal C}} + H.c.]
\label{M_R}
\end{equation}
is not invariant with respect to the $SU(3)_f\times U(1)_s$ symmetry of $\nu_R$:  The field $(\nu_R)^{{\cal C}}$ is a left-handed field
transforming as an anti-triplet, and the product $3^* \times 3^* = 3_a + 6_s^*$ does not contain unity. The Pauli principle
uniquely selects the symmetric sextet. {\bf The global $U(1)_s$ is also evidently broken. Complete spontaneous breakdown of
$SU(3)_f \times U(1)_s$ by $M_f$ is therefore manifest, and it is nicely realized in the Higgs mechanism with sextet \cite{19bhs}.} \\
(ii) The Majorana mass term of $\nu_L$
\begin{equation}
{\cal L}_{M_L} = -\tfrac{1}{2}[\bar \nu_L M' (\nu_L)^{{\cal C}} + H.c.]
\label{M_R}
\end{equation}
is not invariant with respect to the $SU(3)_f$ symmetry of the model: The field $(\nu_L)^{{\cal C}}$ is a right-handed field
transforming as an anti-triplet, and the product $3^* \times 3^* = 3_a + 6_s^*$ does not contain unity.
Moreover, as the left-handed neutrinos belong to the EW doublets the dynamical generation of their Majorana masses is unlikely \cite{20hosek}.\\
(iii) The Dirac mass term of any four SM species $\psi^{(s)}$ $(s = \nu, e, u, d)$
\begin{equation}
{\cal L}_{D} = -[\bar \psi^{(s)}_{R} m \psi^{(s)}_{L} + H.c.]
\end{equation}
is not invariant with respect to the $SU(2)_L \times U(1)_Y$ symmetry of the model: First, the fields $\psi^{(s)}_{R}$ are the EW singlets whereas
$\psi^{(s)}_{L}$ belong to the EW doublets (group-theoretically $1 \times 2 \neq 1+1$). Second, the fixed weak hyper-charges break $U(1)_Y$.
Third, the QED with hard charged-fermion masses remains invariant. In flavor space the Dirac mass belongs both to singlet and octet in
$3^* \times 3= 1 + 8$. Hence, {\bf according to the WS idea the spontaneous EW symmetry breaking pattern by Dirac fermion masses
is identical with that due to the Higgs vacuum condensate}.

The spontaneously generated chiral-symmetry-breaking {\bf soft fermion masses} $M_f$ and $m_f$ are field-theoretically defined
in terms of the chiral-symmetry-breaking fermion self-energy functions $\Sigma_f(p^2)$ in the full fermion propagators
\begin{equation}
S^{-1}_f(p)=\slashed p - \Sigma_f(p^2)
\label{S}
\end{equation}
as their poles: $\mu_f^2\equiv\Sigma_f^2(p^2=\mu_f^2)$. Here $\mu$ abbreviates both $m$ and $M$. The appropriate chiral-symmetry-breaking
$\Sigma_f(p^2)$s are eventually found as the {\bf strong-coupling solutions} of the chirally symmetric Schwinger-Dyson (SD) equations.

With emphasis on the SM sector the paper is organized as follows:\\
In Sect.II we demonstrate the feasibility of spontaneous generation of huge Majorana masses $M_f$ of sterile right-handed neutrinos,
and of three very small Dirac masses $m_f$ of SM fermions by the strong-coupling QFD. We cannot claim for uniqueness,
but the choice of ${\cal L}_f$ required by the WS idea is severely restricted: First, it should be UV safe i.e., asymptotically
free i.e., non-Abelian. Second, to generate also its elementary-fermion mass spectrum the dynamics must not be confining i.e.,
it must be chiral. Third, the new strong force has to be unobservable at present energies i.e., its flavor gluons must be very heavy.
The complete {\it self-consistent} breakdown of $SU(3)_f \times U(1)_s$ by the dynamically generated Majorana masses $M_f$ is,
apparently, unique, and it is in fact responsible for the possibility of revival of the WS idea. Fourth, the extension of the SM fermion sector
by one EW-sterile neutrino $\nu_R$ per flavor {\it by itself} is both innocent and useful. There should be no doubt that Weinberg and Salam
always knew that it would result in neutrino masses. Only for the lack of any experimental evidence for the massiveness of neutrinos
in old days the founders rejected such an extension as unrealistic. It was not their oversight. {\it Gauging} the $SU(3)$ flavor symmetry
with SM fermion sector extended by the right-handed neutrinos with all chiral fermions in flavor triplet makes, however,
the great difference: First, there is the whole interacting BSM sector with one stable very heavy EW-sterile $\nu_R$. Second, in the SM sector
the theoretically safe QFD replaces bona fide the Higgs mechanism, and yields the observable predictions.

At this exploratory stage we leave the elaboration of the consequences of complete dynamical breakdown of $SU(3)_f\times U(1)_s$ symmetry
by $M_f$ in the complex flavor sextet in the form technically already done in \cite{19bhs}, albeit in different context.
The resulting picture is clear and convincing: Twelve real scalar fields of the $\nu_R$-composite sextet decompose into: (i) Eight
'would-be' NG bosons giving rise to huge masses of all eight flavor gluons; (ii) one massive pseudo-NG Majoron of spontaneously broken global
anomalous $U(1)$ symmetry; (iii) three massive Higgs particles above three Majorana neutrino condensates $M_f$.

In sharp contrast with the Higgs mechanism the WS realization of the EW pattern of spontaneous breakdown of $SU(2)_L \times U(1)_Y$ to
$U(1)_{em}$ by three $m_f$ implies: First, there is no genuine Fermi or the EW scale (Sect.III): $m_{W,Z} \sim g(\sum m^2_f)^{1/2}$ and
$\sim ((g^2+g^{'2})^{1/2}(\sum m^2_f)^{1/2}$ respectively i.e., the only scale in the game is the huge scale $\Lambda$. Second, there
should be three SM-fermion-composite Higgs bosons $h_f$ as the small $0^{+}$ excitations above three different SM fermion-anti-fermion
condensates of three fermion-composite doublets defining $m_f$ (Sect.IV):
\begin{widetext}
\begin{eqnarray*}
\phi_f \equiv \tfrac{1}{4v^2}[\bar e_R D_f l_L + (\bar \nu_R D_f l_L)^{\cal C} + \bar d_R D_f q_L + (\bar u_R D_f q_L)^{\cal C}]
\end{eqnarray*}
\end{widetext}
Here $D_1=\tfrac{1}{3}1+\tfrac{1}{2}\lambda_3+\tfrac{1}{2\surd 3}\lambda_8$, $D_2=\tfrac{1}{3}1-\tfrac{1}{2}\lambda_3+\tfrac{1}{2\surd 3}\Lambda_8$,
$D_3=\tfrac{1}{3}1-\tfrac{1}{\surd 3}\lambda_8$ are the diagonal $3\times 3$ matrices $(1, 0, 0), (0 , 1, 0), (0, 0, 1)$ respectively.
The superscript ${\cal C}$ abbreviates the charge conjugation and $v=246 GeV$ is the effective Fermi scale.
The difference with the Higgs mechanism in this case is easily understood:
The EW doublet, both elementary and composite, has only one v.e.v. One condensing elementary EW doublet (together with its charge conjugate)
is enough for phenomenologically parameterizing any number of both upper and lower fermion masses trivially by appropriate Yukawa coupling constants
{\bf provided the new invariant Yukawa interaction with one elementary Higgs doublet (together with its charge conjugate) is postulated}.
The notion of flavor $SU(3)$ symmetry \cite{15flavorchanging}, manifest in perturbation theory, is entirely irrelevant. In the BSM sector the difference
from the Higgs mechanism formally disappears: one composite sextet develops three vacuum expectation values $M_f$ as the elementary scalar sextet;
consequently, there are three BSM $\nu_R$-composite Higgs fields. It is utmost important that, {\it provided the masses $M_f, m_f$ are
dynamically generated as argued}, the above results are practically independent of the badly known functional form of $\Sigma_f(p^2)$.\\

In Sect.V we outline the properties of the weakly coupled EW interactions obtained by including the effects of QFD. We apply to QFD
the dynamical perturbation theory (DPT) of Pagels and Stokar \cite{18ps} originally invented for QCD. It results in the new peculiar $\Sigma_f(p^2)$-dependent
terms in the EW Ward-Takahashi (WT) identities. Consequently, the UV-finite EW results strongly depend
on the functional form of $\Sigma_f(p^2)$. We hesitate to compute the UV-finite loop integrals with poorly known $\Sigma_f(p^2)$, and
make merely the educated expectations. There are, however, the generically new properties:
(i) In contrast with the Higgs mechanism the decays of the composite Higgs bosons into the EW gauge bosons $A,W,Z$ are all fermion-loop
generated. The structure of the phenomenologically important tree-level SM vertices $HWW, HZZ$ is reproduced, but there are the new
effective vertices. Particularly interesting is the description of the manifestly gauge-invariant golden channel
$h_f\rightarrow \gamma \gamma$. (ii) The EW DPT is ideally suited for the UV finite mass splitting between the QFD-degenerate
SM-fermion species in each flavor by the weak hyper-charges. They are thus responsible not only for defining the electric charges
of SM fermions but also for their mass splitting. The dependence of the fermion mass splitting in flavors on $m_f/m_{W,Z}$ accounts for
the fact that the SM fermion mass ordering in flavors is not universal. Unfortunately, the experimental test of this natural idea is not in sight.
The Majorana neutrino mass eigen-states are given in terms of $M_f, m_{f \nu}$ by seasaw. In Sect.VI we summarize,
and briefly discuss the phenomenological usefulness of the theory-enforced BSM extension of the SM by the sector
of EW-sterile QFD-interacting right-handed neutrinos.

We are well aware that the reliable non-numerical understanding of the strong-coupling in QFD (and similarly in QCD) is not in sight.
Consequently, the practical use of the effective field theory underlying QFD, dealing with the 'right' \cite{21wein3}
degrees of freedom (i.e. three interacting composite Higgs bosons $h_f$) should be developed in all details. It is suggestive that it was {\it the effective GL theory,
not BCS, which predicted} in superconductivity two phenomena awarded by Nobel prizes: the Josephson effect \cite{22joseph}, and the Abrikosov's
type-II superconductivity \cite{23abri}.

\section{II. The QFD-generated fermion masses}
First necessary condition for realizing the WS program is the dynamical fermion mass generation by QFD.
It is generally accepted to attribute it to
finding the strong-coupling symmetry-breaking solutions $\Sigma(p^2)$ of the matrix QFD Schwinger-Dyson (SD) equation \cite{24pagels}
\begin{widetext}
\begin{eqnarray}
\Sigma(p)=3\int \frac{d^4k}{(2\pi)^4}\frac{\bar
h^2_{ab}((p-k)^2)}{(p-k)^2}T_a(R)\Sigma(k)[k^2-\Sigma^{+}(k)\Sigma(k)]^{-1}T_b(L)
\label{Sigma}
\end{eqnarray}
\end{widetext}
depicted in Fig.1
\begin{figure}[ht]
\includegraphics[width=0.45\textwidth] {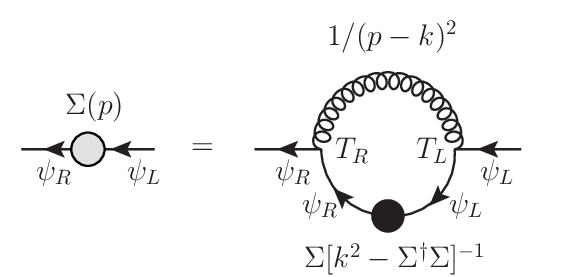}
\caption[99]{The loop-built bridge between the right-handed and the left-handed fermion field.}
\label{obrazek}
\end{figure}
Its explicit form defines both $M_f$ and $m_f$ in the chirally symmetric massless Lagrangian ${\cal L}_f$:
In fact this step, by itself approximate, demands extra model assumptions: In particular, the functional form of the momentum-dependent
coupling matrix $\bar h^2_{ab}((p-k)^2)$ of asymptotically free QFD is explicitly known only
in the non-interesting weak-coupling regime of large momenta.

{\bf The present model is viable if and only if there is a convincing argument that the non-perturbative strong-coupling low-momentum
solutions $\Sigma_f(p^2)$ of the QFD SD equation for this quantity are such that the masses $m_f$ in flavor singlet and octet
come out naturally much much lighter than $M_f$ in the flavor sextet. As all chiral fermions transform as flavor triplets,
the masses $m_f$ come out degenerate for all SM fermion species in given flavor.}

The argument due to Leonard Susskind \cite{25susskind} to which we refer is appealing: In the weakly coupled Higgs mechanism
the dependence of the fermion masses on the Yukawa couplings is linear, and any huge mass difference or ratio is consequently
plainly unnatural: It requires vastly different numbers {\it in the Lagrangian}. In contrast, the strong-coupling {\it non-perturbative}
matrix solutions $\Sigma_f(p^2)$ of the SD equation depend upon the gauge coupling strength $h^2$ multiplied by the corresponding
Clebsch-Gordan coefficients non-analytically. Such a dependence can 'easily' give rise to the desirable huge fermion mass amplification
{\it in solutions of the SD equation}.

To incorporate our ignorance of the strong-coupling regime
phenomenologically we proceed as follows \cite{26hosek}: First, above the scale $\Lambda$ we set $h^2_{ab}((p-k)^2)=0$: In this approximation
QFD is not asymptotically, but exactly free at large momenta. Second, we replace in (\ref{Sigma}) the unknown quantity
$\bar h^2_{ab}((p-k)^2)/(p-k)^2$ by another unknown quantity, the kernel $K_{ab}((p-k)^2)$. It enables to {\it assume}
that the symmetry $SU(3)_f \times U(1)_s$ is dynamically completely broken, and then to find its symmetry-breaking form.

Without loss of generality we fix in the resulting SD equation the external Euclidean momentum as $p=(p,\vec 0)$,
integrate over angles and get
\begin{equation}
\Sigma(p)=\int_0^{\Lambda}k^3dk
K_{ab}(p,k)T_a(R)\Sigma(k)[k^2+\Sigma^{+}\Sigma]^{-1}T_b(L)
\label{Sigmasep}
\end{equation}
Here the {\it unknown} kernel is
\begin{equation}
K_{ab}(p,k)\equiv \frac{3}{4\pi^3}\int_0^{\pi}\frac{\bar
h_{ab}^2(p^2+k^2-2pk \rm \cos \theta)}{p^2+k^2-2pk \rm \cos
\theta}\rm \sin^2 \theta d \rm \theta
\end{equation}
For finding $M_f$ we must set $T_a(R)=\tfrac{1}{2}\lambda_a$ and $T_b(L)=-\tfrac{1}{2}\lambda_b^T$; for finding $m_f$ we must set
$T_a(R)=\tfrac{1}{2}\lambda_a$ and $T_b(L)=\tfrac{1}{2}\lambda_b$.

We have illustrated the Susskind's idea in the separable approximation to the kernel of the SD equation (\ref{Sigmasep})
in the form
\begin{equation}
K_{ab}(p,k)=\frac{3}{4\pi^2} \frac{g_{ab}}{pk}\label{sep}
\end{equation}
in which the SD equation is trivially integrated. The functional form of $\Sigma_f(p^2)$ is of course
\begin{equation}
\Sigma_f(p^2)=\alpha_f (1/p)
\end{equation}
The difficult part is the solution of the homogeneous nonlinear algebraic gap equation for the constant matrix $\alpha_f$.
The point is that with the effective coupling constants $g_f, g'_f$ of the same order of magnitude we get
the enormous mass splitting
\begin{eqnarray*}
M_f &=& g_f \Lambda\\
m_f &=& exp(-1/g'_f) \Lambda
\end{eqnarray*}
In any case the separable approximation (\ref{Sigmasep}) is ideal for trying various Ansatze for $\Sigma_f(p^2)$, and provides a hope
for guessing an appropriate functional form of $\Sigma_f(p^2)$.

To summarize, we proceed with a rather strong assumption: {\bf The spontaneous fermion mass generation by QFD with the property $m_f\ll M_f$
is the reliable solution of the SD equation.} As argued by Susskind the assumption is natural. Moreover, there is the crude model which
supports this assumption. Finally, in the Higgs mechanism (which we replace by QFD), the property $m_f\ll M_f$ is a safe result,
although put by hands \cite{17yana1}.

The use of the WS BCS idea of the non-Abelian chiral strong force with one parameter for dynamical EW symmetry breaking
has the simple and most desirable consequence: The fermion mass spectrum $M_f, m_f$, if dynamically generated by QFD, is
necessarily UV finite and {\it ultimately} calculable \cite{18ps}. The word 'ultimately' is necessary: At present
the reliable strong-coupling tools for the dynamical generation of the fermion masses and their consequences are not known.

{\bf Fortunately, in the SM we are accustomed to a similar situation: In QCD we trust even though the results of its strong-coupling
color-confinement regime are calculable at present also only ultimately}:
In theorists' paradise \cite{27heiri}, the pretty approximation in QCD with massless quarks of the first flavor
and with spontaneously broken chiral symmetry $SU(2)_V \times SU(2)_A$ down to unbroken isospin the hadron masses,
widths of resonances, cross sections, ... are all {\it ultimately} calculable in terms of the only dimensional parameter,
the scale $\Lambda_{QCD}\sim 250 MeV$ of QCD. Unfortunately, the numerical lattice methods successfully used in the
confining vector-like QCD apparently do not apply \cite{28kaplan} in QFD with chirality-prohibited Majorana masses.

\section{III. The QFD-generated $\bf{m_W, m_Z}$ masses}
Second necessary condition for realizing the WS program is the demonstration that the SM fermion self-energies $\Sigma_f$ dynamically generated
by QFD imply by Goldstone theorem the softly massive $W$ and $Z$ bosons. This point is theoretically safe and its demonstration is known \cite{13jjcn}.
It follows the suggestion of Schwinger \cite {29schwinger}: {\it The gauge boson mass squared equals the residue at the massless pole of the transverse
gauge boson polarization tensor.} The derivation of the formulas for $m_W, m_Z$ is available for example in \cite{30bbh},\cite{26hosek}.

For better understanding of the connection of this step with the composite Higgs bosons discussed in Sect.IV we rephrase it here in terms of
the 'classic' idea of technicolor \cite{14tc},\cite{31sikivie}.

First, the {\bf degeneracy of both lepton and quark masses in each flavor} implies that their dynamical generation represents
the explicit realization of spontaneous breakdown of the QFD symmetry $SU(2)_V \times SU(2)_A$ down to $SU(2)_V$ in each flavor, both for quarks and leptons.
Since the currents of $SU(2)_V \times SU(2)_A$ are the known parts of the EW currents we can refer to technicolor \cite{14tc}
and write the general mass formulas for $m_W, m_Z$ in the form
\begin{eqnarray*}
m_W^2 = \tfrac{1}{4}g^2 \sum f_f^2\\
m_Z = \tfrac{1}{4}(g^2+g'^2) \sum f_f^2
\end{eqnarray*}
Here $f_f$, the direct analogs of $f_{\pi}$ in QCD are the vertices of the bilinear derivative couplings of three pseudoscalar
'would-be' NG bosons with three $W,Z$ gauge fields in QFD.

Second, there is the explicit {\it model} of computing $f_{\pi}$ in QCD in terms of the dynamical (constituent) quark masses
in terms of their QCD-generated $\Sigma(p^2)$ by the famous Pagels-Stokar formula \cite{18ps}:
\begin{equation}
f_{\pi}^2=8N\int\frac{d^4p}{(2\pi)^4}\frac{\Sigma^2(p^2)-\tfrac{1}{4}p^2(\Sigma^2(p^2))^{'}}{(p^2+\Sigma^2)^2}
\label{ps}
\end{equation}

Third, we are in fact in the QFD theorists' paradise: Unlike QCD, the spontaneously broken $SU(2)_V \times SU(2)_A$ applies both to leptons and quarks.
Hence, we replace the color factor $N=3$ in the PS formula by $N=1+3=4$. Further, unlike QCD, the $\Sigma_f(p^2)$ generated by QFD describe
the observable fermion masses. Since the form of $\Sigma_f(p^2)$ is currently unknown we write
\begin{equation}
\Sigma_f(p^2)\equiv m_f \sigma(p^2)
\label{smallsigma}
\end{equation}
where $\sigma$ is an unknown universal dimensionless function.
The summation over flavors results in the formulas
\begin{eqnarray*}
m_W^2 &=& \tfrac{1}{4}g^2 c\sum_f m_f^2\\
m_Z^2 &=& \tfrac{1}{4}(g^2 + g'^2) c\sum_f m_f^2
\end{eqnarray*}
where $c$ is an unknown number. An important phenomenological constraint is $c\sum_f m_f^2 = v^2$ where $v  = 246 \rm GeV$.
Hence, $v$ is the {\it effective} Fermi scale. There is no genuine EW mass scale because there is no condensing elementary Higgs field.
The only scale in the model is the QFD scale $\Lambda$. The numerically known quantity $v$ fixes the mass scale of the SM sector of the present model.

\section{IV. The QFD-composite Higgs bosons}
After the 2012 CERN LHC discovery of the Higgs boson $H$ any conceivable realization of the WS BCS-motivated program is obliged
to provide the third necessary condition: The reliable arguments for the existence of the fermion-composite Higgs particle(s).
In the weakly coupled SM the existence of just one Higgs boson is mandatory: Each field in the Lagrangian, apart from the redundant
gauge degrees of freedom, has the unique particle interpretation. In particular, the triplet of the elementary 'would-be' NG bosons
is accompanied by one real scalar Higgs field in one complex Higgs doublet.
In the strongly coupled QFD we can refer merely to the existence theorem of Goldstone, and we are not aware
of any existence theorem for the composite $0^{+}$ Higgs excitation(s).

The natural working hypothesis is to define the fermion-composite Higgs particles as the symmetry partners
of the fermion-composite 'would-be' NG bosons. There is, however, a serious warning: In the confining QCD paradise
so successfully paraphrased as technicolor in Sect.III {\bf there are no symmetry partners of the pions}.
It is a firm experimental fact that the realization of spontaneous chiral $SU(2)_L \times SU(2)_R$ symmetry
breaking down to the isospin symmetry in the effective field theory underlying the confining QCD is
the non-linear $\sigma$-model \cite{21wein3}.

Hence, in the chiral QFD paradise the SM-fermion-composite Higgs bosons as the symmetry partners
of the 'would-be' NG bosons are viable if the realization of spontaneous chiral symmetry breaking
in the effective field theory underlying the chiral QFD is the linear $\sigma$-model. Such a view is in accordance with
the view of Yoichiro Nambu \cite{12nambu}. Fortunately, the present model provides simultaneously an independent, though so far elusive,
test of the hypothesis in its BSM sector: In the $\nu_R$-composite $SU(3)$ sextet there are three composite Higgs $0^{+}$ bosons
in the composite sextet in accord with our hypothesis.

The key point is this: We define the SM-fermion-composite Higgs doublets in the form:
\begin{widetext}
\begin{eqnarray*}
\phi_f \equiv \tfrac{1}{4v^2}\ [\bar e_R D_f l_L + (\bar \nu_R D_f l_L)^{\cal C} + \bar d_R D_f q_L + (\bar u_R D_f q_L)^{\cal C}]
\end{eqnarray*}
Explicitly,
\begin{equation*}
\phi_f \equiv \frac{1}{4v^2}\left[\,
 \left(\begin{array}{c} -\bar l D_f T^{-}\gamma_5 l\\
\bar l D_f l - \bar l D_f T_3 \gamma_5 l\end{array}\right) +
 \left(\begin{array}{c} -\bar q D_f T^{-}\gamma_5 q  \\
	\bar q D_f q - \bar q D_f T_3 \gamma_5 q\end{array}\right)
	\, \right]	\ ,	
\end{equation*}
\end{widetext}
where we have used the known properties of the charge conjugation. The resulting picture is crystal clear:

A. In agreement with Sect.II
\begin{equation*}
<\phi_f>_0 =  m_f
\end{equation*}

B. In agreement with Sect.III the pseudoscalar fermion bilinear combinations describing the 'would-be' NG bosons seen in the WT identities are
tied with each other and with $W,Z$.

C. As far as we can see the scalar $0^{+}$ components of $\phi_f$ are untied and define
three SM-fermion composite Higgs bosons
\begin{equation*}
h_f = \tfrac{1}{4v^2}(\bar l D_f l + \bar q D_f q)
\end{equation*}

D. The effective interaction of the composite Higgses $h_f$ with fermions should have the form
\begin{equation}
{\cal L}_Y = \tfrac{m_f}{v}(\bar l D_f l + \bar q D_f q)h_f
\label{Ly}
\end{equation}
It is the direct partner of the pseudoscalar iso-vector coupling between the pions and quark pseudoscalar densities
in the linear $\sigma$-model \cite{14tc},\cite{31sikivie}.

E. Our definition of the fermion-composite Higgs particles as the small excitations above the
{\bf vacuum condensates} of the spin-zero composite fields guarantees their spin-parity $0^{+}$. This is in accord
with Kibble's general non-Abelian theory of the Higgs mechanism \cite{32kibble}. In this (conservative) sense
the $0^{-}$ or even the charged Higgs bosons, considered in the literature, are an {\it oxymoron}.

We expect that three $h_f$ having the quantum numbers of the vacuum
mix, and undergo the EW corrections. Elaboration of these potentially phenomenologically important steps is outside the scope of the present paper.\\

\section{V. The EW sector influenced by QFD}
Replacement of the weakly coupled Higgs sector by the strong-coupling QFD has for the Standard model the important consequences.
First of all it generates three ($i=1,2,3$) multi-component SM-fermion-composite 'would-be' NG bosons made of $\bar l D_f T_i\gamma_5 l + \bar q D_f T_i\gamma_5 q$
(resulting in the next step in $W,Z$ boson masses), and their three symmetry partners, the SM-fermion-composite composite Higgs bosons $h_f$.
Moreover, the QFD influences the weakly coupled EW interactions as follows:

A. The {\bf fermion propagators} (\ref{S}) acquire the spontaneously QFD-generated chiral-symmetry-breaking $\Sigma_f(p^2)$ i.e.,
the soft fermion masses.

B. There are the effective fermion-Higgs-boson vertices (\ref{Ly}) of the QFD origin proportional to the fermion masses $m_f$.

C. The {\bf  perturbative gauge EW interactions} are uniquely modified by the spontaneously QFD-generated chiral-symmetry-breaking $\Sigma_f(p^2)$
vertices due to the EW Ward Takahashi identities. Being the consequence of the symmetry of the Lagrangian they are non-trivial
namely if the symmetry is spontaneously broken. They have the form
\begin{widetext}
\begin{eqnarray}
\Gamma_A^{\mu}(p',p) &=& eQ_i[\gamma^{\mu} - (p' + p)^{\mu}\Sigma'(p',p)]\label{gammaA}\\
\Gamma_W^{+\mu}(p',p) &=& \tfrac{e}{2 \surd 2 sin\theta_W}\{[\gamma^{\mu} - (p' + p)^{\mu}\Sigma'(p',p)]T^{+}-[\gamma^{\mu}\gamma_5 T^{+}-\tfrac{(p'-p)^\mu}{(p'- p )^2}(\Sigma(p')+\Sigma(p))\gamma_5 T^{+}]\}\label{gammaW}\\
\Gamma_Z^{\mu}(p',p) &=&\tfrac{e}{sin 2 \theta_W}\{[\gamma^{\mu} - (p' + p)^{\mu}\Sigma'(p',p)](T_{3L}^i - 2 Q_i  sin^2 \theta_W)-[\gamma^{\mu}\gamma_5-\tfrac{(p'-p)^{\mu}}{(p'-p)^2}(\Sigma(p')+\Sigma(p))\gamma_5] T_{3L}^i\}
\label{gammaZ}
\end{eqnarray}
\end{widetext}
where
\begin{eqnarray}
\Sigma'(p',p) &\equiv& \frac{\Sigma(p') - \Sigma(p)}{p'^2 - p^2},
\label{sprime}
\end{eqnarray}

Remarks:\\
1. The $\Sigma$-dependent pseudoscalar massless NG poles in(\ref{gammaW}) and in (\ref{gammaZ}), the consequences
of the powerful Goldstone theorem were already employed in Sect.II. Here we emphasize the following: These vertices persist also
in heuristic, unrealistic unrenormalizable NJL models in which $\Sigma(p^2)$ is a constant. In such models, however,
the mass formula for $m_{W,Z}$ cannot be complete: The term with the derivative of $\Sigma$ in the PS formula (\ref{ps}) is simply missing.

2. We will demonstrate in the following that the polar-vector terms in the EW WT identities are equally important:
These new vectorial momentum-dependent vertices without $\gamma$ matrices changing chirality are proportional to the derivatives
of the symmetry-breaking  $\Sigma_f(p^2)$. Hence, their consequences are entirely absent in the NJL models.

\subsection{1. Interactions of $h_f$ with $WW, ZZ$}
In the lowest DPT order the amplitudes of these processes are defined by the fermion-loop triangle Feynman diagrams
with one $\Sigma_f(p^2)$ either in the fermion propagator or with one $\Sigma'_f(p^2)$-dependent vertex. Clearly,
the diagrams are UV-finite. In the first case due the approximate relations $m_W \sim g m_f$, $m_Z \sim (g^2 + g'^2)^{1/2} m_f$
where $m_f$ is the dominant mass the resulting effective Lagrangian for one $h_f$ is of the form of phenomenologically
important tree-level SM Lagrangian
\begin{equation}
{\cal L}_{h_f} \sim (g m_W W^{+}_{\mu}W^{\mu} + \surd (g^2+g'^2) m_Z Z_{\mu}Z^{\mu})h_f
\end{equation}
It is important that the other terms are entirely new and represent the genuine predictions of the present model.
The corresponding computations are postponed for further work.

\subsection{2. The golden channel $h_f\rightarrow \gamma \gamma$}
The exact EM gauge invariance of this channel, crucial for the Higgs-boson identification demands the exact treatment.
The amplitude of this process must result in the transverse form
\begin{equation}
M = F(k_1.k_2 g^{\mu \nu}-k_2^{\mu}k_1^{\nu})\epsilon_{\mu}(k_1) \epsilon_{\nu}(k_2)
\label{M}
\end{equation}
In the Standard model $F$ is the known number which consists of contributions from the separately UV-finite fermion and the $W$ loops
computed most simply in the unitary gauge with hard masses and the canonical SM vertices in that gauge \cite{33twogamma}. Numerically,
it is dominated by the top-quark loop.

In the present model with the SM-fermion-composite Higgses $h_f$ there are no tree-level couplings of $h_f$ with $W,Z$.
Consequently, the decay proceeds solely via the fermion loops. In a paper to be published \cite{34pb} Petr Benes shows that also in this case
the decay amplitude has the requested form (\ref{M}) for arbitrary form of $\Sigma_f(p^2)$.

First, the dynamically generated chirality-changing fermion self-energy $\Sigma(p^2)$ demands the
fermion propagators in the form (\ref{S}). Second, the WT identity for $S^{-1}$ gives rise to the $\Sigma'$-dependent
fermion-fermion-gamma vertex $\Gamma_A^{\mu}$ defined in (\ref{gammaA}).
Third, because there are two photons in the final state the WT identity for $\Gamma_A^{\mu}$
results in the entirely new fermion-fermion-two-photon vertex  $\Gamma^{\mu \nu}(k_1,k_2,p',p)$ which depends also on $\Sigma''$.
The lack of knowledge of the detailed functional form of $\Sigma(p^2)$ is therefore most acute in this case. Hence, again,
the UV finite constant $F$ cannot be computed at present.

\subsection{3. The electroweak SM-fermion mass splitting}

One inspirative point in the prophetic paper \cite{18ps} of Pagels and Stokar is the computation of the UV-finite EM mass difference of
the dynamical (constituent) $u,d$ quarks. The authors use the QCD-generated degenerate soft mass $\Sigma(p^2)$ in the lowest order of their DPT
with the fermion propagator (\ref{S}) and the EM vertex (\ref{gammaA}).

We apply their idea to the QFD-generated soft (current) both lepton and quark masses $\Sigma_f(p^2)$ degenerate in each of three flavors.
These $\Sigma_f(p^2)$ show up in the new vertices of the vectorial EW WT identities (\ref{gammaA},\ref{gammaW},\ref{gammaZ})
and in the full fermion propagators (\ref{S}). {\bf The suggested computation {\it ultimately} results in the UV-finite {\bf electroweak}
SM-fermion mass splitting in all three flavors.}

For different species (s = neutrino, charged lepton $e$, upper quark $u$ with Q=2/3, lower quark $d$ with Q=-1/3) in all three flavors
\begin{equation}
\Sigma_{f s}(p^2)\equiv \Sigma_f(p^2) + \delta_{s}\Sigma_f(p^2)
\label{sigmasf}
\end{equation}
where $\delta_{s}\Sigma_f(p^2)$ is the fermion mass one-loop quantum EW correction in the lowest order of the DPT of Pagels and Stokar
described in Ref.\cite{18ps} and also depicted there. The result is
\begin{widetext}
\begin{eqnarray*}
\delta_s \Sigma_f(p^2) &=& \frac{(eQ_s)^2}{(2\pi)^4}\int \frac{d^4k}{k^2 [(p-k)^2 + m_f^2]}\{3\Sigma_f(p-k) + 2\frac{[\Sigma_f(p)-\Sigma_f(p-k)][p^2k^2-(p.k)^2]}{k^2(k^2-2p.k)}\}\\
                  &+& \frac{(eP_W)^2}{(2\pi)^4}\int \frac{d^4k}{(k^2 + m_W^2)[(p-k)^2 + m_f^2]}\{3\Sigma_f(p-k) + 2\frac{[\Sigma_f(p)-\Sigma_f(p-k)][p^2k^2-(p.k)^2]}{k^2(k^2-2p.k)}\}\\
                  &+& \frac{(eP_Z)^2}{(2\pi)^4} \int \frac{d^4k}{(k^2 + m_Z^2)[(p-k)^2 + m_f^2]}\{3\Sigma_f(p-k) + 2(T_{3L}^s - 2 Q_s  sin^2 \theta_W)^2\frac{[\Sigma_f(p)-\Sigma_f(p-k)][p^2k^2-(p.k)^2]}{k^2(k^2-2p.k)}\}
\end{eqnarray*}
\end{widetext}
where $p,k$ are the Euclidean variables. Here $Q_{\nu}=0, Q_e=-1, Q_u=2/3, Q_d=-1/3$, and for all $s$ $P_W=1/2 \surd 2 sin\theta_W$, $P_Z=1/sin 2\theta_W$.

Remarks:\\
1. The photon propagator is taken in the transverse (Landau) gauge, and the nominators of the $W,Z$ propagators are taken in the transverse
('renormalizable') form.\\
2. Mass splitting due to the $W$ exchange is species-independent in this approximation.\\
3. In denominators of the fermion propagator the hard fermion mass $m_f$ is used for simplicity.\\
4. The EM contribution agrees with the Eq.(17) of Ref.\cite{18ps}).\\
5. It is perhaps worth of mentioning that the symmetry properties of rather unfamiliar $\Sigma'$-dependent fermion-gauge boson vertices
are such that their $\gamma$-matrix-dependent contributions to $\delta_s \Sigma_f(p^2)$ identically vanish (as they should).

With reliable functional strong-coupling form of $\Sigma_f(p^2)$ the integrals in $\delta_s \Sigma_f$ are easily computed.
The solution of the algebraic non-linear equation (with $\Sigma_{f s}(p^2)$
in Minkowski space !)
\begin{equation*}
m^2_{f s} \equiv  \Sigma^2_{f s}(p^2=m^2_{f s})
\end{equation*}
then defines the values of the lepton and quark masses $m_{f s}$ of all SM species $s$ for all three flavors $f$.
Due to the $W,Z$ contributions the fermion masses depend not only on the electric charges,
but also on the ratios $m_f/m_{W,Z}$. This dependence can account for the fact
that the ordering of fermion masses in different flavors is not universal. We do not see any reason why the fermion mass splitting
could not be large as observed.

Regrettably, the price for the new strong force is high: As the reliable form of the strong-coupling $\Sigma_f(p^2)$
is at present unknown the UV-finite computation of the lepton and quark mass spectrum is only 'gedanken'.

\section {VI. Summary and an outlook}
Replacement of the many-parameter weakly interacting Higgs sector of the EW gauge model with the rigid one-parameter strongly coupled QFD
wants to be an attempt towards Weinberg's reductionist vision \cite{1wein1}.
The present picture blurred by the insufficient knowledge of strong coupling is painted as follows:

1. Simple but subtle observation that each SM Dirac-fermion mass breaks the EW $SU(2)_L \times U(1)_Y$ chiral symmetry
exactly down to the observed unbroken $U(1)_{em}$ as the elementary Higgs-field condensate suggests to replace the tree-level Higgs
mechanism by a predictive quantal option. We believe it is in accord with the first Weinberg's law of progress
in theoretical physics which states: "Never believe in the lowest order of perturbation theory" \cite{21wein3}. \\

Present model, similarly to its BCS inspiration, should, first of all, offer the microscopic explanation of the successful
Higgs model, and predict the new observable phenomena. {\bf Provided there are three SM flavors as strongly suggested by LEP
the present model indeed post-dicts the (SM-fermion-composite) $0^{+}$ Higgs boson, and predicts the existence of two new
SM-fermion-composite $0^{+}$ Higgs bosons at the effective Fermi scale $v=246 GeV$ with ultimately calculable properties.}
As the genuine scale $\Lambda$ of QFD is phenomenologically huge the compositeness scale of $h_f$ is apparently unobservably small.

The structure of the SM sector of the present model is given by the pattern of spontaneous breakdown of the EW $SU(2)_L \times U(1)_Y$
symmetry down to $U(1)_{em}$ by three QFD-generated masses $m_f$ degenerate for all SM species in one flavor.
Two microscopic QFD ingredients, the solutions of the SD equations giving rise to $m_f$, and the WT identities
giving rise to three 'would-be' NG poles result in the intuitively transparent
picture of three SM-fermion-composite doublets: (i) Their vacuum condensates are equal to $m_f$. (ii) Their nine pseudoscalar
NG components tied with the gauge fields $W$ and $Z$ by the WT identities into three multi-component bilinear 'would-be' NG bosons are responsible
for their masses proportional to $g(\sum m^2_f)^{1/2}$ and  $(g^2+g{'2})^{1/2}(\sum m^2_f)^{1/2}$, respectively. The degeneracy of
the SM-fermion masses in flavors, direct consequence of the flavor assignment to the chiral fermions,
is utmost important for validity of the observed Weinberg ratio $m_W/m_Z = cos \theta_W$. In sharp contrast
with the SM there is no genuine EW or Fermi mass scale $v=246 GeV$ in the present model. The effective EW scale is the huge mass scale
$\Lambda$ exponentially suppressed {\it in solutions} of the SD equation for the SM-fermion propagators. (iii) As far as we can see
the three scalar components of the composite doublets having the quantum numbers of the vacuum are untied, and remain in the spectrum
as the $0^{+}$ SM-fermion-composite Higgs bosons $h_f$ with fixed effective Yukawa couplings. Their expected mixing is not
considered at this exploratory stage. (iv) The appearance of the composite Higgs bosons apparently does not generate any hierarchy problem \cite{35peskin}.
(v) In contrast with the elementary SM Higgs boson $H$ all couplings of the composite Higgs bosons $h_f$
with the EW gauge bosons $A, W, Z$ exhibit explicitly the electroweak unification i.e., they are all treated on the same footing:
They are all generated by the UV-finite SM-fermion loops.

Since the microscopic strong-coupling QFD solution, similarly to QCD, is not in sight, the practical use of the underlying effective field theory
dealing with three interacting composite Higgs bosons $h_f$ should be developed in all details.
{\it In the effective GL theory, not in BCS} two phenomena awarded by Nobel prizes, the Josephson effect \cite{22joseph},
and the Abrikosov's type-II superconductivity, were predicted \cite{23abri}. The second Weinberg's law of progress
in theoretical physics states: "You can choose any degrees of freedom for the description of your system. But if you use
the wrong ones you will be sorry." \cite{21wein3}

Optimistic conclusion is that the Higgs boson $H$ discovered at the CERN LHC in 2012 is the first of three
composite Higgses $h_f$ suggested here. The modest indications from the CERN LHC (and from LEP)
of the existence of extra Higgses do exist \cite{36newh}. Perhaps we may say that the large experimental uncertainties
in fixing the properties of $H$, and the large theoretical uncertainties in fixing the properties of $h_f$ leave this point at present inconclusive.\\

Because of the lack of the reliable functional form of $\Sigma_f(p^2)$ the Sect.V. is rather brief. It is intended mainly to illustrate
the {\bf potential} predictability of the present model.

Tiny effects of interactions of flavor gluons with masses of order $\Lambda$ with the SM fermions
were neglected at the present stage for simplicity. In old days before the discovery of $W,Z$ the small departures
from the Fermi's four-fermion weak interaction indicated the existence of the $W,Z$ bosons of the SM by the propagator
effects; similarly, there could be the four-fermion terms marking the tiny departures from the SM. Because the detailed
elaboration of the BSM sector is postponed to future work, we did not make any serious attempt at fixing $\Lambda$.
Here we just mention that its lower limit is apparently given by the absence of flavor-changing neutral currents.
Various uses of sterile right-handed neutrinos with masses also of order $\Lambda$ discussed here and in the literature suggest,
however, its much much higher values.

2. As implicitly contained already in the original BCS WS idea the present BSM with the strong-coupling QFD provides
a theoretically consistent framework for the calculable fermion mass spectrum:

In the first step the strong $SU(3)_f$ QFD spontaneously distinguishes three fermion flavors by three different huge
{\bf soft} Majorana masses $M_f$ of sterile $\nu_{fR}$, and by three different small {\bf soft}  Dirac masses $m_f \ll M_f$
{\it common to all SM fermion species in $f$}. Both $M_f$ and $m_f$, strictly prohibited by the chiral symmetries of the model
together with the corresponding mass counter-terms must be, if dynamically generated, the finite and calculable multiples
of $\Lambda$ \cite{18ps}. Similarly to the 'theorists' QCD paradise' \cite{27heiri} {\it these six dimensionless numbers
are the inherent property of the chiral $SU(3)_f$ QFD}.

In the second step the mass degeneracy of the SM fermion species in flavors is {\it naturally split by the weak hyper-charges}
in the UV-finite weakly coupled {\bf electroweak} DPT of Pagels and Stokar. If the word 'natural' is appropriate somewhere
it is here. The originally theoretically arbitrary weak hyper-charges are undoubtedly the ideally suited parameters for this purpose.
Not only that they are fixed by the observed SM fermion electric charges; they at the same time determine also their masses.
Moreover, the dimensionless ratios $m_f/m_{Z,W}$ entering the computation are well suited for accounting the experimental fact
that the fermion mass ordering in flavors is not universal.
Finding the reliable $\Sigma_f$ apparently requires a new bright idea. Merely "churning the equations gets nowhere."
(The third Weinberg's law of progress in theoretical physics \cite{21wein3}) Fifty years of experience with QCD teaches us to be patient.
In any case the postulate of the SM Yukawa $H$ and $H^{\cal C}$ interactions with SM-fermions and with arbitrary coupling constants
becomes superfluous. For computation of the mass spectrum of six Majorana neutrinos there is the third necessary step,
{\it the famous seasaw} \cite{37seasaw} with ultimately calculable $M_f$ and $m_{f \nu}$.

3. The structure of the BSM sector of the present model follows closely the structure of the SM sector:\\
It is fixed by the pattern of the complete spontaneous breakdown of the $SU(3)_f\times U(1)_s$ symmetry of the $\nu_{fR}$ sector
by three QFD-generated sterile neutrino masses $M_f$ contained in the $\nu_{fR}$-composite sextet. The resulting picture is quite transparent
(12=5+3+3+1): (i) Five pseudoscalar and three scalar 'would-be' NG bosons in the sextet give rise to the huge masses
of all flavor gluons proportional to $M_f$ i.e., to the currently unobservable short-range QFD. (ii) There are three super-heavy
sterile-neutrino-composite $0^{+}$ vacuum-like Higgs-like bosons $\chi_i$. Honestly, we are not aware of any BSM phenomenon which
would really call for them. (iii) The sterile-neutrino-composite pseudo-NG Majoron $\eta_M$ of spontaneously broken global anomalous
$U(1)_s$ sterility symmetry is different: It is the theoretically reliable and phenomenologically welcome prediction of the model:
It can be identified with a particle-physics-natural inflaton \cite{39freese}.
(iv) Finding the spectrum of the strongly coupled $\nu_{fR}$ composites is an important task for the future. (v) In any case the present model
contains all ingredients needed in popular BSM extensions: The seasaw \cite{37seasaw}, the elegant, conservative baryogenesis without GUTs \cite{38yana},
and the inflation \cite{39freese}. Closeness of the huge mass scales of all these phenomena and our $\Lambda$ is particularly suggestive.
The detailed elaboration of the BSM sector of the present model deserves the separate dedicated work.\\

4. The present model emphasizes the all-important role of fermions and their masses. Here we come to their many-body, macroscopic role.

The many-electron BCS derives, in accord with observations, the exponential behavior of the specific heat
of the macroscopic quantum super-conducting state \cite{40fw}. Its shape is entirely due to the gap in the energy spectrum of the
Bogolubov-Valatin (BV) \cite{41bogo, 42vala} quasi-electrons, the true fermion excitations above the BCS ground state. Specific heat
of the low-temperature Landau Fermi liquid, the many-body system of dressed electrons above the critical super-conducting temperature
$T_c$ is observed, again in accord with observations, as linear \cite{40fw}.

What is the many-body role of fermions in the SM and in the BSM? The spontaneous formation
of the gap in the spectrum of the BV quasi-electrons due to the Cooper-pair vacuum condensate $<a^{+}_{\vec k \uparrow}a^{+}_{-\vec k \downarrow}>_0$
of the electrons $a_{\vec k \sigma}$ violating the particle number conservation is to be compared with the formation of the fermion-antifermion
vacuum condensates $<\bar \psi_R \psi_L>_0$ and $<\bar \nu_R (\nu_R)^{{\cal C}}>_0$ i.e., the fermion masses spontaneously violating the chiral symmetry.
Consequently, the superconductivity analogy does not suggest any sort of superconductivity or superfluidity in our Lorentz-invariant world.

The many-body role of fermions of the SM sector with their dynamically generated masses of leptons and quarks is in giving the mass to
the stable luminous Universe. Due to the confining QCD its mass is formed by the stable systems made of electrons and quarks of the first flavor
in which the free-neutron decay is strictly prohibited by the powerful Pauli principle. Similarly, we expect that there are the stable
systems of the QFD-formed $\nu_{fR}$-composites in the BSM sector. If indeed formed, for us they would be the natural candidate
for the dark matter of the Universe.

5. Finally, the theoretically arbitrary scale $\Lambda$ of the present model is phenomenologically necessarily huge.
It might thus be relevant to mention that the {\it theoretical need of a huge particle-physics mass scale} is the necessary condition
of Andrei Sakharov's effective-field-theory concept of the emergent gravity \cite{43sakh}. The clear modern picture of such a concept
connecting particle physics with gravity is provided by Steven Adler \cite{44adler}. The consequences are far-reaching.
For representative examples we refer to \cite{45adler}, \cite{46zee}, \cite{47staro}.

I am grateful to Petr Bene\v s for permission to present his results prior to publication.

\end{document}